\documentclass[aps,
prx,
superscriptaddress,
amsmath,
amssymb,
reprint,
twocolumn,
floatfix]
{revtex4-2}

\usepackage[utf8]{inputenc}
\usepackage{graphicx}
\usepackage{color}
\usepackage{qcircuit}
\usepackage{braket}
\usepackage{url}
\usepackage{comment}
\usepackage{bm}
\usepackage[version=4]{mhchem}
\usepackage{mathtools}
\usepackage{siunitx}
\usepackage[breaklinks=true]{hyperref}
\newcommand{\mr}[1]{\mathrm{#1}}

\date{\today}

\begin{document}
\title{Quantum Car-Parrinello Molecular Dynamics: A Cost-Efficient Molecular Simulation Method on Near-Term Quantum Computers}

\author{Kohdai Kuroiwa}
\email{kuroiwa@qunasys.com}
\affiliation{QunaSys Inc., Aqua Hakusan Building 9F, 1-13-7 Hakusan, Bunkyo, Tokyo 113-0001, Japan}
\author{Takahiro Ohkuma}
\email{takahiro.ohkuma@bridgestone.com}
\author{Hirokazu Sato}
\email{hirokazu.sato@bridgestone.com}
\affiliation{Digital Engineering Division, Bridgestone Corporation, Kodaira, Tokyo 187-8531, Japan}
\author{Ryosuke Imai}
\email{imai@qunasys.com}
\affiliation{QunaSys Inc., Aqua Hakusan Building 9F, 1-13-7 Hakusan, Bunkyo, Tokyo 113-0001, Japan}

\begin{abstract}
In this paper, we propose a cost-reduced method for 
finite-temperature 
molecular dynamics on a near-term quantum computer, \textit{Quantum Car-Parrinello molecular dynamics} (QCPMD). One of the most promising applications of near-term quantum computers is quantum chemistry. 
It has been expected that simulations of molecules via molecular dynamics can be also efficiently performed on near-term quantum computers by applying a promising near-term quantum algorithm of the variational quantum eigensolver (VQE).
However, this method may demand considerable computational costs to achieve a sufficient accuracy, and otherwise, statistical noise can significantly affect the results. 
To resolve these problems, we invent an efficient method for molecular time evolution inspired by Car-Parrinello method. In our method, parameters characterizing the quantum state evolve based on equations of motion instead of being optimized. 
Furthermore, by considering Langevin dynamics, we can make use of the intrinsic statistical noise. 
As an application of QCPMD, we propose an efficient method for vibrational frequency analysis of molecules in which we can use the results of the molecular dynamics calculated by QCPMD. 
Numerical experiments show that our method can precisely simulate the Langevin dynamics at the equilibrium state, and we can successfully predict a given molecule's eigen frequencies. 
Furthermore, in the numerical simulation, our method achieves a substantial cost reduction compared with molecular dynamics using the VQE. 
Our method achieves an efficient computation without using widely employed method of the VQE. In this sense, we open up a new possibility of molecular dynamics on near-term quantum computers. 
We expect our results inspire further invention of efficient near-term quantum algorithms for simulation of molecules. 
\end{abstract}
\maketitle

\section{\label{sec:Intro} Introduction}

\textit{Ab-initio} molecular dynamics (AIMD)~\cite{Hutter2012,Tuckerman2002,Iftimie2005,Paquet2018,Carloni2002} is a powerful technique to analyze properties of chemical molecules and condensed matter systems, such as chemical reaction processes~\cite{Mosconi2015,Zhang2015}, diffusion properties~\cite{He2018,Baktash2020,Stamminger2020},
liquid~\cite{Hassanali2011,Zen2015}, perovskites~\cite{Johari2011,Zhang2016}, amorphous materials~\cite{Mattoni2016}, and vibrational frequencies~\cite{Zhang2011,Thomas2013}. 
Despite its advantages, however, AIMD requires to solve the eigenvalue problem to obtain the eigenstates of the electrons in a given molecule repeatedly, which demands enormous computational resources.

Quantum computers are one of the most expected candidates to make a breakthrough against this challenge. 
In particular, with recent advancement of quantum technologies, near-term quantum computers called noisy intermediate-scale quantum (NISQ) devices~\cite{Preskill2018} have been getting attraction as practically realizable quantum computers in the coming near future.

A NISQ device is a quantum computer that possesses less than a few hundred qubits and does not perform error correction on its qubits.   
While NISQ devices are not suitable to perform complicated quantum algorithms, such as quantum phase estimation~\cite{Kitaev1995,Cleve1998} and quantum searching~\cite{Grover1996}, that require a large number of qubits and error correction, simulating quantum many-body systems is one of the most promising ways to utilize NISQ devices. 

Indeed, it was proposed to demonstrate AIMD on NISQ devices~\cite{Fedorov2020,Sokolov2021}.
In these approaches, one uses NISQ devices to estimate the nuclear gradients at each time step of the dynamics. 
In addition, a classical-quantum hybrid algorithm called the \textit{variational quantum eigensolver} (VQE)~\cite{Peruzzo2014} is used for the eigenvalue problem to obtain the electron ground state. 
The VQE was already performed on real NISQ devices~\cite{Peruzzo2014,Kandala2017,Colless2018}, and it is expected to solve the eigenvalue problem more efficiently than classical computers for a large quantum system, for example, chemical molecules~\cite{Mcardle2018,Cao2018}.

Even with the high expectation, however, there are still serious problems upon the performance of AIMD on NISQ devices.
First, the statistical noise in the computation is inevitable in principle because we can only acquire results by sampling with a finite number of samples from quantum computers. 
Sokolov \textit{et al.}~\cite{Sokolov2021} proposed to make use of the statistical noise in evaluation of force to simulate a finite-temperature molecular dynamics via generalized Langevin dynamics. 
Nevertheless, the successful optimization by the VQE is required, which is hard to guarantee if the number of samples is limited.
Second, the VQE still demands many samples if we wish to simulate a large molecule~\cite{Wecker2015} even though cost reduction in the VQE has been widely studied~\cite{Wecker2015,ZhouLeo2020,Nakanishi2020,Parrish2019,Kubler2020,Arrasmith2020,Lee2019,Gard2019,DallaireDemers2019,Ibe2020,Wiersema2020,McClean2016,Hamamura2020, Jena2019,Verteletskyi2020,Yen2020,Izmaylov2020,Gokhale2020,Bravyi2017,WangDaochen2019}. (See Sec.~\ref{subsec:VQE_cost} for more details.) 
Because AIMD involves time evolution of large molecules, we need to run the VQE algorithm as a subroutine repeatedly in the procedure of AIMD. 
Since we have to require that the VQE yield sufficiently accurate results at each step of AIMD, 
a further technical breakthrough is demanded to efficiently perform AIMD on NISQ devices.
Thus, it is not convincingly hopeful that we can practically demonstrate AIMD using the VQE algorithm. 

In this paper, we propose a method to simulate AIMD on NISQ devices, which we call \textit{Quantum Car-Parrinello molecular dynamics} (QCPMD).
Our idea is inspired by the method of Car-Parrinello molecular dynamics (CPMD)~\cite{Car1985}, in which the wave-functions are introduced as fictitious dynamical variables and the electronic ground state is approximately computed by considering a virtual dynamics of the electronic wave-functions. 
Instead of finding the ground state of a given electron Hamiltonian at each time step using the VQE, we update the parameters in an \textit{ansatz state} according to an artificial equations of motion and let the nuclei and the parameters evolve in parallel.
A simulation with QCPMD is expected to be less costly than that using the VQE because we do not require the parametrized molecular electronic state to converge to the ground state at each time step. 
We formulate the parameter update as a Langevin dynamics as well to make use of statistical noise emerging from evaluation of observables in time evolution of parameters. 
With this method, we can describe the motions of the nuclei and the parameter updates in a unified way. 

We first show the procedure of QCPMD and provide its theoretical analysis and justification. 
By simultaneously considering a Langevin dynamics of the nuclei and parameters, 
our method simulates dynamics in a constant temperature under the equilibrium probability distribution.
Then, as an application of QCPMD, we also propose a way to analyze vibrational frequencies of molecules from the results of the molecular dynamics. 
Our method is based on an approximation of the Hessian matrix, which is needed to calculate the frequencies of a molecule, proposed in Ref.~\cite{Luo2014}. 
While a direct evaluation of a Hessian matrix may become highly costly, we successfully simplifies the calculation, which is applicable with the results of QCPMD. 
Finally, we perform numerical simulations of QCPMD for the $\ce{H_2}$ molecule. 
We analyze the vibrational modes for these molecules using the method we proposed. Our results show that the simulation by QCPMD yields sufficiently accurate results to predict the dominant vibrational modes with improved computational costs.

Hence, our method of QCPMD offers a way to making use of intrinsic statistical noise of quantum computers. Moreover, it also provides a unified perspective for updates of the nuclei configuration and the parameters of a electronic state in a sense that they evolve according to equations of the same form. 
In addition, our method of QCPMD successfully reduces computational costs for AIMD compared with the previous method. 
Furthermore, using the results of QCPMD to approximate the Hessian matrix, we can correctly estimate molecular vibrational frequencies. 
Our method achieves a key cost reduction, opening up a way to apply near-term quantum computers to AIMD.

The rest of this paper is organized as follows. 
In Sec.~\ref{subsec:MDVQE}, we explain the common setup of AIMD using near-term quantum computers. 
In Sec.~\ref{sec:Quantum CP}, we introduce our new method of QCPMD, and discuss the theory of this method. 
In Sec.~\ref{sec:applications}, we propose a way to apply QCPMD to molecular vibrational frequency analysis. We further conduct numerical simulations to verify QCPMD performs efficiently and accurately compared to the previous methods. 
Finally, in Sec.~\ref{sec:conclusion}, we summarize our results and discuss the future direction. 

\section{\label{subsec:MDVQE} Review for Setups of Molecular Dynamics on Near-Term Quantum Computers}
In this section, we briefly review an approach of molecular dynamics using near-term quantum computers. 

AIMD consists of two main components: electronic states and nuclei configuration. 
In AIMD, we treat the nuclei of given molecules as classical mass points. That is, we consider that the nuclei move according to the potential energy surfaces obtained by calculating the ground-state energy of the electron subsystem. 
We adopt the Born-Oppenheimer approximation~\cite{Born1927}, which assumes that the motions of the atomic nuclei of given molecules are much slower than those of the electrons due to the fact that the nuclei are far heavier than the electron. 
By this approximation, we can treat the electron dynamics and nuclear dynamics separately.
Then, as long as the Born-Oppenheimer approximation is valid, AIMD can successfully predict the behaviour of the given molecules. 

In Sec.~\ref{subsec:estimation_ground_state}, we review the estimation method of the electronic ground state on NISQ devices. 
In Sec.~\ref{subsec:estimation_force} we discuss the evaluation of nuclei force, which is needed to update nuclei configuration. 
In Sec.~\ref{subsec:procedure_of_AIMD}, we explain the procedure of AIMD on a quantum computers that uses the VQE algorithm. 

Hereafter in this paper, we consider the dynamics of $N$ molecules.  
Let $H(\bm{R})$ be the Hamiltonian of the given molecules with $\bm{R} \coloneqq (\bm{R}_1,\ldots,\bm{R}_N)$ denoting the nuclear coordinates where $\bm{R}_k \coloneqq (R_{k,x}, R_{k,y}, R_{k,z})$ is the coordinate of the $k$th nucleus.
In addition, for $N$-dimensional vectors $\bm{a}$ and $\bm{b}$, we define
\begin{align}
    \bm{a}\bm{b} &\coloneqq (a_kb_k)_{k=1,\ldots,N}\\
    \bm{a}^{n} &\coloneqq (a_k^n)_{k=1,\ldots,N}\\
    \frac{\bm{a}}{\bm{b}} &\coloneqq \left(\frac{a_k}{b_k}\right)_{k=1,\ldots,N}\\
    \sqrt{\bm{a}} &\coloneqq (\sqrt{a_k})_{k=1,\ldots,N}
\end{align}
when every entry of the right-hand sides are well-defined.

\subsection{Estimation of Electronic State}~\label{subsec:estimation_ground_state}
Here, we overview a procedure to calculate the electronic state of the Hamiltonian $H(\bm{R})$ using quantum computers. 
For this purpose, we consider the second quantized formulation of the Hamiltonian. 
We transform the second quantized Hamiltonian into a qubit Hamiltonian using some mapping method, \textit{e.g.,} Jordan-Wigner transformation~\cite{Jordan1928} and Bravyi-Kitaev transformation~\cite{Bravyi2002,Seeley2012}.
For more details, see also Refs.~\cite{Mcardle2018,Cao2018}. 
After we have a qubit Hamiltonian $H(\bm{R})$, we represent the electronic state of the molecules by an ansatz state $\ket{\psi(\bm{\theta})}$, which is characterized by parameters $\bm{\theta} = (\theta_1,\ldots,\theta_M)$. Then, we update the parameters so that the ansatz state will be close to the ground state of Hamiltonian $H(\bm{R})$. 
For example, the VQE algorithm (see Sec.~\ref{subsec:VQE}) is expected to yield a good approximation of the ground state. 
With Hamiltonian $H(\bm{R})$ and electronic state $\ket{\psi(\bm{\theta})}$, 
\begin{equation}~\label{eq:energy_eval}
    L(\bm{\theta},\bm{R}) \coloneqq \braket{\psi(\bm{\theta})|H(\bm{R})|\psi(\bm{\theta})}
\end{equation}
is an approximate ground-state energy.  

\subsection{Estimation of Forces}~\label{subsec:estimation_force}
Here, we show the way we estimate forces on the nuclei. 
Once we obtain the energy $L(\bm{\theta},\bm{R})$ as shown in the previous section, we calculate the forces $\bm{F}(\bm{\theta},\bm{R}) \coloneqq (\bm{F}_1(\bm{\theta},\bm{R}), \ldots, \bm{F}_N(\bm{\theta},\bm{R}))$ on the nuclei. 
For each $k \in \{1,2,\ldots,N\}$, the force $\bm{F}_k(\bm{\theta},\bm{R}) \coloneqq (F_{k,x}(\bm{\theta},\bm{R}),F_{k,y}(\bm{\theta},\bm{R}),F_{k,z}(\bm{\theta},\bm{R}))$ on the $k$th nucleus is given by 
\begin{equation}
    F_{k,\alpha}(\bm{\theta},\bm{R}) = -\frac{\partial}{\partial R_{k,\alpha}} L(\bm{\theta},\bm{R}) 
\end{equation}
for $\alpha = x,y,z$. 
The right-hand side can be expressed as  
\begin{widetext}
\begin{equation}
    \begin{aligned}
    \frac{\partial}{\partial R_{k,\alpha}} L(\bm{\theta},\bm{R})
    &= \Braket{\psi(\bm{\theta})\left|\frac{\mathrm{d} H(\bm{R})}{\mathrm{d}_{k,\alpha}}\right|\psi(\bm{\theta})} 
    + \Braket{\partial_{R_{k,\alpha}}\psi(\bm{\theta})| H(\bm{R}) |\psi(\bm{\theta})}
    +\Braket{\psi(\bm{\theta})| H(\bm{R}) |\partial_{R_{k,\alpha}}\psi(\bm{\theta})},
    \end{aligned}
\end{equation}
\end{widetext}
where 
\begin{equation}~\label{eq:partial_R}
    \ket{\partial_{R_{k,\alpha}}\psi(\bm{\theta})} \coloneqq \frac{\partial}{\partial R_{k,\alpha}}\ket{\psi(\bm{\theta})}.
\end{equation}
The first term is called the Hellman-Feynman force~\cite{Hellmann1937,Feynman1939}, and the sum of the second and third terms are called the Puley force~\cite{Pulay1969}. 
If $\ket{\psi(\bm{\theta})}$ is the exact ground state, the Puley forces vanish~\cite{Guttinger1932,Pauli1933,Hellmann1937,Feynman1939}. 
One only considers the Hellman-Feynman force assuming that $\ket{\psi(\bm{\theta})}$ is sufficiently close to the ground state. 
The Hellman-Feynman force is obtained by estimating the expectation value of the gradient of the Hamiltonian.
The gradient can be derived analytically using the results of the Pauli measurements~\cite{Yamaguchi1994,Mitarai2020,OBrien2019}. 
If we already performed the Pauli measurements in the evaluation of the Hamiltonian given in Eq.~\eqref{eq:energy_eval}, we can reuse these results in the evaluation of the gradient instead of conducting measurements again. 

Note that, practically, when we evaluate the force, we cannot ignore effects of statistical noise due to limited number of samples. 
The noise in the force leads to non-conservation of the mechanical energy. 

\subsection{Procedure of AIMD}~\label{subsec:procedure_of_AIMD}
We show the procedure of AIMD on quantum computers using the estimation of electronic states and nuclei forces introduced in the previous subsections. 
Here, we adopt the VQE algorithm to estimate the electronic ground state. 

Suppose that we are given a molecule consisting of atoms with mass $\bm{m}$ and initial nuclei positions $\bm{R}_0$. We are also given an ansatz state $\ket{\psi(\bm{\theta})}$.
Then, by using the VQE algorithm, we have the optimal parameters $\bm{\theta}^*(\bm{R}_0)$ for this nuclei positions, for which $\ket{\psi(\bm{\theta}^*(\bm{R}_0))}$ is a good approximation of the ground state. 
With this approximate ground state, we can estimate nuclei forces $\bm{F}(\bm{\theta}^*(\bm{R}_0),\bm{R}_0)$ by using the method shown in Sec.~\ref{subsec:estimation_force}. 
We numerically update the positions of the nuclei according to the equations of motions $\bm{m}\ddot{\bm{R}}_0 = \bm{F}(\bm{\theta}^*(\bm{R}_0),\bm{R}_0)$ where parameters $\bm{\theta}^*(\bm{R}_0)$ are fixed.
Letting $\bm{R}_1$ denoting the updated nuclei positions, we perform the same procedure again for $\bm{R}_1$, and get an approximate ground state for $\bm{R}_1$ and updated positions $\bm{R}_2$. 
By repeating this procedure, we obtain the nuclei positions $\bm{R}_n$ and an approximate ground state $\ket{\psi(\bm{\theta}^*(\bm{R}_n))}$ at the $n$th step; that is, we can describe the time evolution of the molecules.

\section{Quantum Car–Parrinello Molecular Dynamics\label{sec:Quantum CP}}
In this section, we show our formulation for \textit{ab initio} molecular dynamics (AIMD) on near-term quantum computers, which we call \textit{Quantum Car-Parrinello molecular dynamics} (QCPMD). 

The main idea of QCPMD is that we treat the nuclei positions $\bm{R}$ and the parameters $\bm{\theta}$ equivalently; that is, we consider they are virtually both mass points moving according to the potential energy surfaces $L(\bm{\theta},\bm{R})$. 
We also adopt Langevin formulation to make use of the effects of statistical noise, which inevitably originates from a finite number of samples. 
Indeed, application of Langevin dynamics in the context of molecular dynamics was proposed in previous research~\cite{Sokolov2021,Luo2014,Attaccalite2008,Thomas2007,Mazzola2014,Mazzola2012,Mazzola2017,Mazzola2018}. 
In particular, Sokolov \textit{et al.}~\cite{Sokolov2021} proposed that Langevin dynamics can be considered in molecular simulations using the VQE algorithm on near-term quantum computers. 
Here, we show an alternative method to apply Langevin dynamics on near-term quantum computers, which may yield a better sampling cost that that using the VQE. 

Motivated by Car-Parrinello molecular dynamics~\cite{Car1985}, we consider updating nuclei configurations and parameters separately by forces
\begin{align}
    \bm{F}(\bm{\theta},\bm{R}) &= -\nabla_{\bm{R}} L(\bm{\theta},\bm{R}), \\ 
    \bm{F_{\theta}}(\bm{\theta},\bm{R}) &= -\nabla_{\bm{\theta}} L(\bm{\theta},\bm{R}),  
\end{align}
where $\bm{F}$ is an actual force on the nuclei and $\bm{F_{\theta}}$ is a virtual force on the parameters. 
In a simulation on a quantum computers, we cannot ignore the effects of statistical noise in evaluating the values of $\bm{F}$ and $\bm{F}_\theta$. 
Let $\tilde{\bm{F}}$ and $\tilde{\bm{F}}_{\bm{\theta}}$ denote the noisy forces, 
and let $\bm{F}$ and $\bm{F_{\theta}}$ denote the true values.
Assuming that the noise can be modeled by a normally distributed random variable, we may write
\begin{align}
    \label{F_noise}
    \tilde{\bm{F}} &= \bm{F} + \bm{f}\bm{W}\\
    \label{F_theta_noise}
    \tilde{\bm{F}}_{\bm{\theta}} &= \bm{F}_{\bm{\theta}} + \bm{f}_{\bm{\theta}}\bm{W}_{\bm{\theta}}.  
\end{align}
In these equations, $\bm{f}$ and $\bm{f}_{\theta}$ are the standard deviations of $\bm{\tilde{F}}$ and  $\bm{\tilde{L}_{\theta}}$ defined as 
\begin{align}
    f_{i} &= \sqrt{\braket{(\tilde{F}_i-\braket{\tilde{F}_i}_{\mathrm{stat}})^2}_{\mathrm{stat}}},\\
    (f_{\theta})_{i} 
    &= \sqrt{\braket{(\tilde{{F_{\theta}}})_i-\braket{({\tilde{{F_{\theta}}})_i}}_{\mathrm{stat}})^2}_{\mathrm{stat}}},  
\end{align}
respectively, 
where $\bm{W}(t)$ and $\bm{W}_{\bm{\theta}}(t)$ describe Gaussian processes following the standard Gaussian distribution. 
In these definitions, $\braket{\cdots}_{\mathrm{stat}}$ is the statistical average over sampling with respect to the given state. 
More precisely, we define $\braket{\cdots}_{\mathrm{stat}}$ as follows. 
Let $O$ be an observable to measure. Then, the outcome $O'$ of the measurement follows some probability distribution $p(O')$ determined by the state of the system and the measurement method. 
The statistical average of $O$ for this measurement is defined as 
\begin{equation}
    \braket{O}_{\mathrm{stat}} = \sum_{O'} p(O')O'. 
\end{equation}

We can model the motion of molecules with this statistical noise using Langevin-type equations by introducing dissipation terms. We may obtain the molecular dynamics under an equilibrium distribution from the simulation of the Langevin-type equations of motion. 
For this purpose, we introduce dissipation terms $\gamma\bm{v}$ and $\zeta\bm{\xi}$ where $\gamma$ and $\zeta$ are friction matrices and $\bm{v} \coloneqq \dot{\bm{R}}$ and $\bm{\xi} \coloneqq \dot{\bm{\theta}}$. 
If a molecule consists of $N$ atoms and parameters $\bm{\theta}$ have $M$ elements, $\gamma$ is a real $3N \times 3N$ matrix and $\zeta$ is a real $M\times M$ matrix. 
Matrices $\gamma$ and $\zeta$ are determined so that they can realize some desired finite temperature. We will show the actual expressions of $\gamma$ and $\zeta$ later in Eqs.~\eqref{eq:gamma_f} and \eqref{eq:zeta_f}, by which we can compute $\gamma$ and $\zeta$ using data obtained from the simulation. 

Let $\bm{R}^{(k)}$, $\bm{v}^{(k)}$, $\bm{\theta}^{(k)}$, and $\bm{\xi}^{(k)}$ denote the values at the $k$th step of the simulation.
Introducing the dissipation terms, equations of motion we compute are
\begin{align}
    \bm{v}^{(k+1)} 
    \nonumber
    &= (1-\gamma(\bm{\theta}^{(k)},\bm{R}^{(k)})\Delta t) \bm{v}^{(k)} \\ 
    \label{disc_v}
    &+ \frac{\bm{F}(\bm{\theta}^{(k)},\bm{R}^{(k)})}{\bm{m}}\Delta t, \\
    \label{disc_R}
    \bm{R}^{(k+1)} 
    &= \bm{R}^{(k)}+\bm{v}^{(k)}\Delta t,\\
    \bm{\xi}^{(k+1)}
    \nonumber
    &= (1-\zeta(\bm{\theta}^{(k)},\bm{R}^{( k+1)})\Delta t) \bm{\xi}^{(k)} \\
    \label{disc_xi}
    &+ \frac{\bm{F_{\theta}}(\bm{\theta}^{(k)},\bm{R}^{(k+1)})}{\bm{\mu}}\Delta t, \\
    \label{disc_theta}
    \bm{\theta}^{(k+1)} 
    &= \bm{\theta}^{(k)} + \bm{\xi}^{(k)}\Delta t,
\end{align}
where $\bm{\mu}$ is the virtual mass of parameters and $\Delta t$ is the time step of the simulations. 
Now, considering statistical errors originating from a finite number of samples, from Eqs.~\eqref{F_noise} and \eqref{F_theta_noise}, we obtain 
\begin{align}
    &\bm{v}^{(k+1)} = (1-\gamma^{(k)}\Delta t) \bm{v}^{(k)} + \frac{\bm{F}^{(k)}}{\bm{m}}\Delta t + \frac{\bm{f}\bm{W}^{(k)}}{\bm{m}}\Delta t, \\
    &\bm{\xi}^{(k+1)}= (1-\zeta^{(k)}\Delta t) \bm{\xi}^{(k)}
   + \frac{\bm{F_{\theta}}^{(k)}}{\bm{\mu}}\Delta t + \frac{\bm{f}_{\theta}\bm{W}_{\theta}^{(k)}}{\bm{\mu}}\Delta t, 
\end{align}
with $\gamma^{(k)} \coloneqq \gamma(\bm{\theta}^{(k)},\bm{R}^{(k)})$ and $\zeta^{(k)} \coloneqq \zeta(\bm{\theta}^{(k)},\bm{R}^{(k+1)})$. 
Standard deviations $\bm{f}$ and $\bm{f}_{\theta}$ are related to the number of samples; 
in particular, it is expected that the standard deviations are inversely proportional to square root of the number of samples for sufficiently large sample size.
That is, letting $\rho$ be the number of samples used to estimate the forces in a unit time,
\begin{align}
    \bm{f} &= \sqrt{\frac{\bm{c}}{\rho\Delta t}} \\
    \bm{f}_{\theta} &= \sqrt{\frac{\bm{c}_{\theta}}{\rho\Delta t}} 
\end{align}
with some positive constants $\bm{c}$ and $\bm{c}_{\theta}$. 
Then, using these relations, the equations of motion become
\begin{align}
    \label{eq:langevin_discrete_v}
    &\bm{v}^{(k+1)} = (1-\gamma^{(k)}\Delta t) \bm{v}^{(k)} + \frac{\bm{F}^{(k)}}{\bm{m}}\Delta t + \sqrt{\frac{\bm{c}}{\rho}}\frac{\bm{W}^{(k)}}{\bm{m}}\sqrt{\Delta t}, \\
    &\bm{\xi}^{(k+1)}= (1-\zeta^{(k)}\Delta t) \bm{\xi}^{(k)} + \frac{\bm{F_{\theta}}^{(k)}}{\bm{\mu}}\Delta t + \sqrt{\frac{\bm{c}_{\theta}}{\rho}}\frac{\bm{W}_{\theta}^{(k)}}{\bm{\mu}}\sqrt{\Delta t}.
\end{align}
Taking the limit of $\Delta t$ being extremely small, we have the following Langevin-type equations:
\begin{align}
    \label{Langevin_v}
    &\mathrm{d}\bm{v} = \left(-\gamma \bm{v}  + \frac{\bm{F}}{\bm{m}}\right)\mathrm{d}t + \sqrt{\frac{\bm{c}}{\rho}}\frac{\mathrm{d}\bm{W}(t)}{\bm{m}}, \\
    \label{Langevin_R}
    &\mathrm{d}\bm{R} = \bm{v}\mathrm{d}t\\
    \label{Langevin_xi}
    &\mathrm{d}\bm{\xi}= \left(-\zeta \bm{\xi} + \frac{\bm{F_{\theta}}}{\bm{\mu}}\right)\mathrm{d}t + \sqrt{\frac{\bm{c}_{\theta}}{\rho}}\frac{\mathrm{d}\bm{W}_{\theta}(t)}{\bm{\mu}}\\
    \label{Langevin_theta}
    &\mathrm{d}\bm{\theta} = \bm{\xi}\mathrm{d}t.
\end{align}
When obtaining Eq.~\eqref{Langevin_v} from Eq.~\eqref{eq:langevin_discrete_v}, $W^{(k)} \sqrt{\Delta t}$ is identified with $dW(t) \coloneqq B_{t+dt} - B_{t}$ as a stochastic process, where $B_t$ is the Winner process with time step $\Delta t$. 
We also employ the same identification of $W^{(k)}_{\theta} \sqrt{\Delta t}$ and $dW_{\theta}(t)$ in Eq.~\eqref{Langevin_xi}. 
Assume that the equilibrium distribution is given by the following Boltzmann distribution up to some normalization constant:
\begin{equation}~\label{Eq:equib_dist}
    \begin{aligned}
    &\Phi(\bm{R},\bm{v};\bm{\theta},\bm{\xi}) \\
    &\propto \exp\left(-\beta \left(\frac{1}{2}\bm{m}\cdot\bm{v}^2 + L(\bm{\theta,\bm{R}})+\frac{1}{2}\bm{\mu}\cdot\bm{\xi}^2\right)\right)
    \end{aligned}
\end{equation}
with some inverse temperature $\beta = 1/(k_BT)$. 
With this equilibrium distribution, we can compute dissipation coefficients $\gamma$ and $\zeta$ using $\bm{f}$ and $\bm{f}_{\bm{\theta}}$. 
Indeed, we have the following fluctuation-dissipation theorem:
\begin{align}
    \label{eq:FD_gamma}
    \gamma_{ij} &= \frac{\beta c_i}{2\rho m_i}\delta_{ij},\\
    \label{eq:FD_zeta}
    \zeta_{ij} &= \frac{\beta (c_{\theta})_i}{2\rho \mu_i}\delta_{ij}.
\end{align}
See Appendix~\ref{sec:FDT} for detailed derivation. 
Note that the off diagonal elements of $\gamma$ and $\zeta$ are zero because the evaluation of each entry of $\bm{F}$ and $\bm{F_{\theta}}$ is independent of each other. 
In our simulation, since $\bm{c}/\rho = \bm{f}^2\Delta t$ and $\bm{c}_{\theta}/\rho = \bm{f}_{\theta}^2\Delta t$, it holds that
\begin{align}
    \label{eq:gamma_f}
    \bm{\gamma}^{\mathrm{diag}} &= \frac{\bm{f}^2\beta\Delta t}{2\bm{m}}\\
    \label{eq:zeta_f}
    \bm{\zeta}^{\mathrm{diag}} &= \frac{\bm{f}_{\theta}^2\beta\Delta t}{2\bm{\mu}}, 
\end{align}
where $\bm{\gamma}^{\mathrm{diag}}$ and $\bm{\zeta}^{\mathrm{diag}}$ are vectors corresponding to the diagonal elements of $\bm{\gamma}$ and $\bm{\zeta}$ respectively. 
Thus, $\gamma$ and $\zeta$ can be evaluated by estimating the statistical variance of the forces. 
Combining Eqs.~\eqref{disc_v}, ~\eqref{disc_R}, ~\eqref{disc_xi}, and ~\eqref{disc_theta} with Eqs.~\eqref{eq:gamma_f} and ~\eqref{eq:zeta_f}, we can simulate the Langevin equations of motions of the given molecule under a preset finite temperature $\beta$. 

Considering the time evolution of these Langevin-type equations of motion, we can simulate the dynamics of given molecules under the equilibrium distribution~\eqref{Eq:equib_dist} with temperature $\beta$; that is, we have sequences of $\{\bm{R}_n\}_{n}$ and $\{\bm{\theta}_n\}_n$ at the $n$th time step. 
Practically, in a simulation of molecular dynamics, we solve discrete equations of motion shown in Eqs.~\eqref{disc_v}-\eqref{disc_theta}.
Since $c$ and $c_{\theta}$ depend on the variables $\bm{R}$ and $\bm{\theta}$, the noise terms of these stochastic differential equations are multiplicative. 
In Eq.~\eqref{disc_xi}, we have used $\bm{F_{\theta}}$ as a function of $\bm{R}^{(k+1)}$ instead of $\bm{R}^{(k)}$. 
It should be noted that in the limit of $\Delta t \to 0$, the difference vanishes because $\bm{R}^{(k+1)} - \bm{R}^{(k)}$ is of the order of $\Delta t$. 
If we consider an overdamped case, the stochastic equation should be interpreted with the Storatonovich representation to reproduce the Boltzmann distribution. 
Herein, we only study the underdamped cases as described in Eqs.~\eqref{Langevin_v}-\eqref{Langevin_theta}.
In each step of the simulation, after we obtain the force, we update the nuclei configuration and the parameters for electronic states according to these stochastic differential equations obtained above. 
In a simulation, the time evolution is computed by numerically integrate Eqs.~\eqref{disc_v}-\eqref{disc_theta} with Euler-Maruyama method~\cite{Maruyama1955}. 

Lastly, we make a justification for using the equilibrium distribution in Eq.~\eqref{Eq:equib_dist} for finite-temperature molecular dynamics. 
One may notice that Eq.~\eqref{Eq:equib_dist} depends on parameters $\bm{\theta}$ and $\bm{\xi}$ while statistics of interest in molecular dynamics mainly involves only $\bm{R}$ and $\bm{v}$. 
When statistics of interest do not depend on $\bm{\xi}$, we can consider the marginal distribution 
\begin{equation}
    \begin{aligned}
    \propto\exp\left(-\beta\left(\frac{1}{2}\bm{m}\cdot\bm{v}^2 + L(\bm{\theta},\bm{R})\right)\right)
    \end{aligned}
\end{equation}
because $\bm{\xi}$ is independent of the other variables. 
Now, the marginal distribution with respect to $\bm{R}$ and $\bm{v}$ is 
\begin{equation}~\label{eq:marginal_dist_int}
    \propto \int \exp\left(-\beta\left(\frac{1}{2}\bm{m}\cdot\bm{v}^2 + L(\bm{\theta},\bm{R})\right)\right)d\bm{\theta}. 
\end{equation}
When $\beta$ is sufficiently large, applying the saddle point approximation~\cite{bender2013}, we have 
\begin{equation}
    \begin{aligned}
    &\sqrt{\frac{(2\pi)^M}{\beta^M |\mathrm{det}B(\bm{\theta}^*(\bm{R}),\bm{R})|}} \\ 
    &\times \exp\left(-\beta\left(\frac{1}{2}\bm{m}\cdot\bm{v}^2+L(\bm{\theta}^*(\bm{R}),\bm{R})\right)\right), 
    \end{aligned}
\end{equation}
where $\bm{\theta}^*(\bm{R})$ is an optimum achieving the minimum value at $\bm{R}$ 
and $B(\bm{\theta}^*(\bm{R}),\bm{R})$ is the Hessian matrix of $L(\bm{\theta},\bm{R})$ with respect to $\bm{\theta}$ at $\bm{\theta}^*(\bm{R})$. 
Hereafter, we assume that $B(\bm{\theta}^*(\bm{R}),\bm{R})$ is almost constant for all $\bm{R}$ the given molecule achieves during the simulation. 
This assumption is, for example, satisfied when $\bm{R}$ does not change largely in the time evolution. 
We will see in Sec.~\ref{subsec:numerical_simulation} that our simulation satisfies this assumption while it is not the case in general, depending on what ansatz circuit we choose.
Thus, as long as we set the temperature not too large in the simulation, we have 
\begin{equation}~\label{eq:marginal_dist}
    \exp\left(-\beta\left(\frac{1}{2}\bm{m}\cdot\bm{v}^2+L(\bm{\theta}^*(\bm{R}),\bm{R})\right)\right)
\end{equation}
up to an appropriate normalization constant. 
This is a Boltzmann distribution in the form of $\exp(-\beta E_{\mathrm{total}}(\bm{R}))$ with the total mechanical energy 
\begin{equation}
    E_{\mathrm{total}}(\bm{R}) \coloneqq \frac{1}{2}\bm{m}\cdot\bm{v}^2+L(\bm{\theta}^*(\bm{R}),\bm{R}).
\end{equation}
Therefore, to compute statistics only depending on $\bm{R}$ and $\bm{v}$, which is usually on main interest, we can apply the marginal distribution~\eqref{eq:marginal_dist} as an effective probability distribution. 

\section{Application of Quantum Car-Parrinello Molecular Dynamics to Vibrational Frequency Analysis}~\label{sec:applications}
In this section, we discuss applications of QCPMD. 
We propose a method to efficiently perform vibrational frequency analysis of a given molecule using results of QCPMD. 
In Sec.~\ref{subsec:vibrational_analysis}, we explain the theory and the procedure of our vibrational frequency analysis method based on the evaluation of Hessian matrix using the results of QCPMD. 
We also compare computational costs of our method with the previous method. 
Moreover, in Sec.~\ref{subsec:numerical_simulation}, we conduct a numerical simulation of QCPMD. The results show that QCPMD efficiently yields accurate results compared with the AIMD based on the naive application of the VQE. 

\subsection{Vibrational Frequency Analysis through Covariance Matrix of Force}~\label{subsec:vibrational_analysis}

\begin{table*}[ht]
    \centering
    \caption{Comparison of methods for estimating Hessian.}
    \begin{tabular}{lcc}
    \hline\hline
     & \multicolumn{2}{c}{Method} \\
     & Direct & QCPMD-Based \\
    \hline
    Require Geometry Optimization?& Yes & No \\
    Require Optimal Parameter?& Yes & No \\
    Require Time-Series Data?& No & Yes \\
    Require High Accuracy Data From Quantum Computer?& Yes & No \\
    \hline\hline
    \end{tabular}
    \label{tab:comparison}
\end{table*}

We show a method of molecular vibrational frequency analysis using the results of QCPMD.
Our idea and implementation are motivated by those in Ref.~\cite{Luo2014}, which gives methods of vibrational frequency analysis of \textit{ab initio} molecular dynamics (AIMD) with noisy forces by quantum Monte Carlo method. 

Before explaining the procedure of our method, we explain key components of molecular vibrational frequency analysis.
We consider (mass-weighted) internal coordinate $\bm{\bar{s}}$ of the given molecules. 
Let $L$ be the cost function, and let $\bm{\bar{s}^*}$ be the equilibrium point such that 
\begin{equation}
    \frac{\partial L}{\partial \bar{s}^{*}_i} = 0
\end{equation}
for all $i$. 
Define the displacement $\bm{s}$ in the internal coordinate as 
$\bm{s} \coloneqq \bm{\bar{s}} - \bm{\bar{s}^{*}}$. 
Considering the harmonic approximation, the motion of the nuclei is given as 
\begin{equation}
    \frac{\mathrm{d}^2}{\mathrm{d}t^2} \bm{s} = -A \bm{s}
\end{equation}
around the equilibrium point, where $A$ is the Hessian defined as 
\begin{equation}
    A_{ij} = \frac{\partial^2L}{\partial s_i \partial s_j}. 
\end{equation}
The eigen frequencies of the given molecule are the eigenvalues of $A$.
Therefore, it is vital to have the Hessian $A$ to conduct vibrational frequency analysis of the molecule. 
However, it may be costly to estimate the Hessian matrix $A$ on near-term quantum computers because it demands $O(N^2)$ calculations of second derivatives. 

Now, we propose a method to estimate $A$ using the results of QCPMD. 
In the simulation, the statistic $\bm{s}$ is distributed with Boltzmann distribution 
\begin{equation}~\label{eq:int_prob_dist}
    \Phi(\bm{s}) \propto \exp\left(-\beta L\right) 
    \approx \exp\left(-\frac{1}{2} \beta \bm{s}^\mathrm{T}  A \bm{s}\right)
\end{equation}
up to some normalization constant. 
Let $C_{\bm{s}}$ is the covariance matrix of the internal coordinate $\bm{s}$ defined as 
\begin{equation}~\label{eq:def_cov_s}
    (C_{\bm{s}})_{ij} = \braket{(s_i - \braket{s_i})(s_j - \braket{s_j})}
\end{equation}
where $\braket{\cdots}$ is the ensemble average defined as 
\begin{equation}
    \braket{f(\bm{s})} \coloneqq \int d\bm{s} f(\bm{s}) \Phi(\bm{s}).
\end{equation}
Since $\braket{\cdots}$ is the average with respect to the probability distribution~\eqref{eq:int_prob_dist}, 
by evaluating Eq.~\eqref{eq:def_cov_s} explicitly using the probability distribution, we have that 
\begin{equation}
    C_{\bm{s}} = \frac{A^{-1}}{\beta}, 
\end{equation}
which is equivalent to 
\begin{equation}~\label{eq:A_Cs}
    A = \frac{C_{\bm{s}}^{-1}}{\beta}. 
\end{equation}
Here, we may assume that the time average agrees with the ensemble average. 
Let us define 
\begin{equation}~\label{eq:covariance_est_time}
    (C_{\bm{s}}^{(\mathrm{exp})})_{ij}
    = \braket{(s_i - \braket{s_i}_{\mathrm{time}})(s_j - \braket{s_j}_{\mathrm{time}})}_{\mathrm{time}}, 
\end{equation}
where $\braket{\cdots}_{\mathrm{time}}$ is the time average; that is, 
\begin{equation}
    \braket{\bm{s}}_{\mathrm{time}} \coloneqq \frac{1}{K}\sum_{k=1}^K \bm{s}^{(k)}, 
\end{equation}
where $\bm{s}^{(k)}$ is the internal coordinates at the $k$th step. 
We may consider $C_{\bm{s}}^{(\mathrm{exp})}$ is an good approximation of $C_{\bm{s}}$ defined in Eq.~\eqref{eq:A_Cs}.
Thus, in a practical setup, we can use the time-series data of coordinates from QCPMD to compute $C_{\bm{s}}$ with Eqs.~\eqref{eq:A_Cs} and \eqref{eq:covariance_est_time}. 
Although Eq.~\eqref{eq:A_Cs} involves the inverse of $C_{\bm{s}}$, we may avoid the calculation of the inverse. 
What is of interest is the eigen frequencies and the normal modes of the given molecules; hence, we only need the eigenvalues and the eigenvectors of $A$. 
to obtain the eigenvalues, we can first compute the eigenvalues of $C_{\bm{s}}$ and take the inverse of the eigenvalues. 
The eigenvectors of a normal matrix and its inverse are the same; we can directly employ the eigenvectors of $C_{\bm{s}}$ as the normal modes.

Now, we estimate the computational cost of this method in terms of the number of evaluations of expectation values. Suppose that we run QCPMD of a given molecules consisting of $N$ atoms $L_{\mathrm{QCP}}$ steps. Let $N_{\theta}$ denote the number of parameters of the ansatz state. At each step of the simulation, we compute the force and update the parameters; then, we need to evaluate expectation values $3N$ distinct operators for the computation of the force and $N_{\theta}$ times for the update of the parameters. In each evaluation, we need $O(1/\epsilon_{\mathrm{QCP}}^2)$ samples to achieve error $\epsilon_{\mathrm{QCP}}$. Therefore, the total cost throughout the simulations is 
\begin{equation}    O\left(\frac{1}{\epsilon_{\mathrm{QCP}}^2}\right) \times (3N + N_{\theta}) \times L_{\mathrm{QCP}}.
\end{equation}

On the other hand, to compute the Hessian in the method directly using the VQE, we first need to accurately obtain the equilibrium geometry of a given molecule. While an approximate geometry can be obtained by using density functional theory methods~\cite{koch2015}, post-Hartree-Fock methods~\cite{jensen2017} may well be needed for more accurate estimation. Recently, a near-term quantum algorithm for the geometry optimization has also been proposed~\cite{Delgado2021}.
After we have an accurate estimation of the equilibrium geometry, we need a good approximation of the optimal parameters by VQE. 
Considering the VQE algorithm with $L_{\mathrm{VQE}}$ optimization steps, we need $O(1/\epsilon_{\mathrm{VQE}}^2)\times L_{\mathrm{VQE}}$ samples to achieve error $\epsilon_{\mathrm{VQE}}$. 
To directly compute the Hessian matrix, we need to evaluate expectation values $O(N^2)$ times if we numerically evaluate the derivative of the Hamiltonian. In each evaluation, we need $O(1/\epsilon^2)$ samples to achieve the accuracy $\epsilon$. 
Therefore, the total cost throughout the computation is 
\begin{equation}
    O\left(\frac{1}{\epsilon_{\mathrm{VQE}}^2}\right) \times L_{\mathrm{VQE}} +  O\left(\frac{1}{\epsilon^2}\right) \times O(N^2)
\end{equation}
even assuming that we already have the equilibrium geometry. 

Hence, our QCPMD-based method for evaluation of Hessian may well be more cost-efficient than the previous direct estimation method. 
Indeed, we observe that the cost of QCPMD-based method is $O(N)$ and while that of the direct estimation is $O(N^2)$. 
In addition, in the direct estimation method, we need to run VQE and to evaluate Hessian highly accurately compared to QCP; that is, $\epsilon_{\mathrm{VQE}}, \epsilon \ll \epsilon_{\mathrm{QCP}}$ because we can take advantage of the error in the QCP procedure. 
We summarize the comparison in Table~\ref{tab:comparison}.

\subsection{Numerical Simulations of Molecular Quantum Car–Parrinello Dynamics}~\label{subsec:numerical_simulation}
In this section, we numerically simulate QCPMD for $\ce{H_2}$ molecule to investigate its performance, \textit{e.g.,} accuracy and computational costs, compared with the molecular dynamics using the VQE (VQE-based MD). 

\subsubsection{Setup of Numerical Simulations}
Our setup for numerical simulations is as follows.
We consider a hydrogen molecule $\ce{H_2}$.
The initial structure is set to the equilibrium nuclear distance and the velocity is zero initially.  
We simulate QCPMD and the molecular dynamics using the VQE by emulating the output of quantum computers by classical computers. 
We adopt the STO-3G minimal basis set. 
We prepare the fermionic second-quantized Hamiltonians for electrons using PySCF~\cite{Sun2018} and Openfermion~\cite{Mcclean2017} based on the structures at each time. We use Jordan-Wigner transformation~\cite{Jordan1928} to map the fermionic Hamiltonians into qubit Hamiltonians~\cite{Mcardle2018, Cao2018}.
We take the time steps $\Delta t = \SI{0.01}{fs}$, and we run the time evolution of the $\ce{H_2}$ molecule for $\SI{4,000}{fs}$ under temperature $70$ K.
We use the real-valued symmetry-preserving type ansatz of the depth $D=4$ introduced in Refs.~\cite{Ibe2020,Gard2019} to perform the molecular dynamics. 
We take $\mu = 0.01$ as the virtual mass of the ansatz parameters for the QCPMD time evolution. 
We employ BFGS method implemented in SciPy for VQE parameter optimization.
We perform measurement on each Pauli operator individually while \textit{grouping} strategies have been proposed, in which we measure multiple Pauli operators simultaneously to reduce the total number of measurements~\cite{Kandala2017}.
We do not include any noise for quantum circuit simulations except for the shot noise.
All simulations are performed by the high-speed quantum circuit simulator Qulacs~\cite{Suzuki2020}.

\subsubsection{Simulations of Hydrogen Molecule}

\begin{figure*}
    \centering
    \includegraphics[width = 0.8\linewidth]{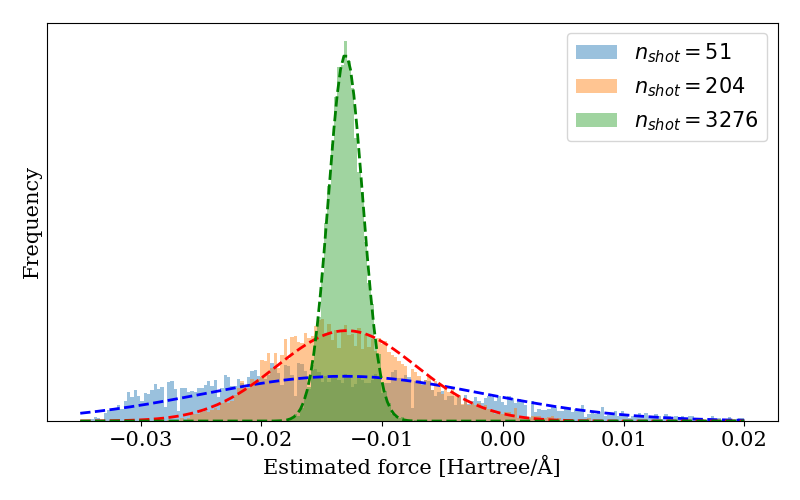}
    \caption{
    Histogram of the distribution of the nuclei force at the initial geometry of the hydrogen molecule for several shot numbers. 
    In this figure, $n_\mathrm{shot}$ is the number of sample used to estimate the expectation value of a Pauli operator.
    We also plot the best Gaussian fitting of each histogram. 
    The dashed blue, red, and green curves are the best Gaussian curves for $n_{\mathrm{shot}} = 51, 204, 3276$ respectively. 
    For the evaluation of Pauli operators, we perform measurement on each Pauli operator individually; that is, we did not adopt grouping~\cite{McClean2016} of Pauli operators. 
    }
    \label{fig:sample}
\end{figure*}

\begin{figure*}
    \centering
    \includegraphics[width =0.8\linewidth]{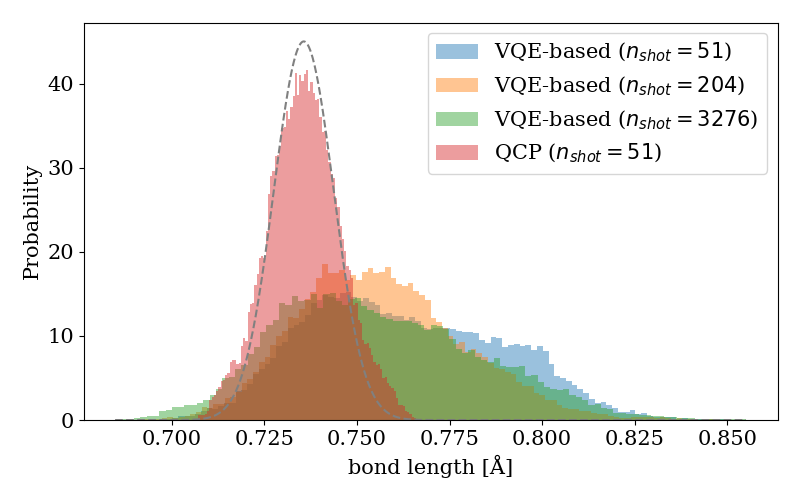}
    \caption{
    The histograms of bond length produced by the VQE and the QCP method. 
    The setup of the simulations are shown at the beginning of Sec.~\ref{subsec:numerical_simulation}. 
    Here, $n_\mathrm{shot}$ is the number of sample used to estimate the expectation value of a Pauli operator.
    }
    \label{fig:r_hist_H2}
\end{figure*}

We show the results of the simulations of the molecular dynamics of $\ce{H_2}$. 

We first justify the modeling of the statistical noise using Gaussian white noise. 
Next, we compare the performance of QCPMD with the method using the VQE. 
To justify that we treat the shot noise as a Gaussian process, we plot the distribution of force at the initial geometry of the given $\ce{H_2}$ molecule in Figure~\ref{fig:sample}. 
We also plot the best Gaussian curve of the distribution of force for each shot number. 
The graph indicates that the force is distributed with a Gaussian distribution for sufficiently large shot numbers. 
Indeed, we observe that the total variational distance between the computed distribution and the best fit Gaussian distribution is $\approx 0.02$.
When we increase the number of samples, while the expectation value is not changed, the variance becomes smaller; that is, the distribution becomes sharp as the number of shots gets large. 
This behavior indicates that the noise is alleviated as the sample size gets large. 

Now, we compare QCPMD and VQE-based MD. 
Our main question is whether QCPMD outperforms VQE-based MD. 
Figure~\ref{fig:r_hist_H2} shows the histogram of the bond length obtained by VQE-based MD and QCPMD.
We see that the equilibrium Gaussian distribution is obtained by QCPMD, which agree with the reference Gaussian distribution predicted from the given computational conditions. 
When using VQE, the correct equilibrium distribution is not obtained even with more than six times as many shots as QCPMD.

We conduct vibrational frequency analysis based on the results of MD. Since VQE-based MD does not yield the correct equilibrium Gaussian distribution as shown in Fig.~\ref{fig:r_hist_H2}, we cannot apply our vibrational frequency analysis method shown in Sec.~\ref{subsec:vibrational_analysis} in this case. 
On the contrary, we can perform the vibrational frequency analysis with QCPMD, for which the results are distributed with a Gaussian distribution. 
The result of the vibrational frequency analysis is shown in Table~\ref{tab:frequency}. 
Here, we assume that the equilibrium (Gaussian) distribution is not achieved in the earlier stage of the simulation, and we discard the first $500\,\mathrm{fs}$ to evaluate the covariance matrix in Eq.~\eqref{eq:covariance_est_time}. 
We adopt the jackknife resampling method~\cite{efron1982} with five bins of the same size for the computation of the frequency to alleviate the bias of the time average. 
We compare the result by QCPMD with the reference value obtained by the full configuration interaction (FCI) method.
We can see that QCPMD gives the correct vibrational frequency of this molecule with a small deviation from the reference.

Lastly, the comparison of computing resources required to perform the Langevin molecular dynamics is shown in Table ~\ref{tab:resource}.
One can observe that QCPMD reduces both the number of shots and circuits compared to VQE method. 
Indeed, the cost of QCPMD is around $1/8$ of that of the VQE-based MD even if $n_{\mathrm{shot}}$ is fixed to $n_{\mathrm{shot}} = 51$ without considering the accuracy. 
From these results and observations, QCPMD outperforms over the VQE-based MD in terms of both accuracy and computational cost. 

\begin{table}[htbp]
    \centering
    \caption{Vibrational frequency analysis for the given $\ce{H_2}$ molecule. In the table, we show the result obtained by QCPMD using the method shown in Sec.~\ref{subsec:vibrational_analysis} and the result obtained by the full configuration interaction (FCI) method as a reference. }
    \begin{tabular}{lcccccccc}
    \hline\hline
     &  QCP ($n_\mathrm{shot} = 51$) & FCI \\
    \hline
    Frequency
     & $4956\,\mathrm{cm}^{-1}$ &
     $4989\,\mathrm{cm}^{-1}$\\
    \hline\hline
    \end{tabular}
    \label{tab:frequency}
\end{table}

\begin{table*}[htbp]
    \centering
    \caption{Computational resources required to perform the Langevin molecular dynamics of the hydrogen molecule per step ($\Delta t = 0.01 \mathrm{fs}$, 400,000 steps). In this table, $n_\mathrm{shot,total}$ is the total number of shots and $n_\mathrm{circuit}$ is the number of circuits used in the entire dynamics.}
    \begin{tabular}{lcccccccc}
    \hline\hline
     & VQE ($n_\mathrm{shot} = 51$) & VQE ($n_\mathrm{shot} = 204$) & VQE ($n_\mathrm{shot} = 3276$) & QCP ($n_\mathrm{shot} = 51$) \\
    \hline
    $n_\mathrm{shot, total}$
     & 252,687 & 1,011,585& 16,309,576 & 37,230 \\
    $n_\mathrm{circuit}$
     & 4,955 & 4,959 & 4,979 & 730 \\
    \hline\hline
    \end{tabular}
    \label{tab:resource}
\end{table*}

\section{Conclusion}~\label{sec:conclusion}
In this paper, we proposed a novel molecular simulation method of QCPMD. 
In this method, we update the nuclei configuration and the parameters of an ansatz state describing the molecule's electron state separately. 
And thus, we does not require the exact ground state but the parameters fluctuate around the optimal values during the simulation with an appropriate choice of fictitious mass $\mu$, that is, sufficiently small $\mu$. 
This ensures that at each step of the simulation, the parametrized ansatz state is close to the ground state of the molecule, leading to the successful simulation of the time evolution. 
Moreover, the time evolution of QCPMD is formulated by using Langevin dynamics, in which we make use of the intrinsic noise to obtain molecular dynamics under the equilibrium distribution. 
While usage of Langevin dynamics for the nuclei configuration was also discussed in the previous research~\cite{Sokolov2021}, QCPMD applies Langevin dynamics to not only the time evolution of nuclei but also the parameter update. 
In this way, we can also take the statistical noise in the parameter update into account. 

As an application of QCPMD, we proposed an efficient way for vibrational frequency analysis of molecules. 
In our method, one no longer needs to evaluate the Hessian matrix at the equilibrium structure, but can directly use results of QCPMD to estimate the Hessian matrix. 
Our method is efficient in the sense that we do not require exact and accurate optimization in QCPMD. 
The numerical simulation we conducted strengthened the justification on the application of QCPMD; it shows that QCPMD performed better than the VQE-based MD with respect to the precision and cost. 

Thus, our method suggested a new direction of applications of near-term quantum computers, in which we do not require the VQE algorithm.
Notwithstanding, although molecular dynamics is a well-established simulation method to analyze properties of molecules, we may not be able to directly apply near-term quantum computers to it considering its cost. 
We showed a way to alleviate this problem and provide an efficient way to perform molecular dynamics on quantum computers, 
which may be a key to the real application of NISQ devices in the near future. 
As a future direction, it is interesting to look for a better molecular time-evolution method based on the idea of QCPMD. 
At the same time, it is also interesting to perform simulations with QCPMD of a larger molecule that cannot be fully explained with the Hartree-Fock method. 
Simulation of QCPMD on a real NISQ device is also interesting direction to verify our method is practically useful in a real sense. 
Our method paves a way for further applications of near-term quantum computers without the VQE, broadening their possibility towards practical applications. 

\section*{Acknowledgments}
The authors are grateful to W.~Mizukami and S.~Koh for valuable discussions during an early stage of this study.
The authors also thank Y.~O.~Nakagawa for his valuable comments on the manuscript.
KK is supported by a Mike and Ophelia Lazaridis Fellowship, the Funai Foundation, and a Perimeter Residency Doctoral Award. 

\bibliography{bibliography}
\newpage 

\appendix
\section{\label{sec:Prelim} Variational Quantum Eigensolver (VQE)}
In this section, we overview the variational quantum eigensolver (VQE), which is one of the most promising classical-quantum hybrid algorithm for near-term quantum computers. 
In Sec.~\ref{subsec:VQE}, we review the basic setup of the VQE algorithm
In addition, in Sec.~\ref{subsec:VQE_cost}, we summarize previous approaches and results to suppress the cost of the VQE. 
\subsection{\label{subsec:VQE} Setup of VQE}
The VQE~\cite{Peruzzo2014} is a quantum-classical hybrid algorithm to compute the ground state and its energy of a given system, which is considered to be executable on NISQ devices. 
Suppose that we have a quantum system with a Hamiltonian $\hat{H}$. We prepare an ansatz state of $n$ qubits,
\begin{equation}
    \ket{\psi(\bm{\theta)}} \coloneqq U(\bm{\theta})\ket{\psi_{0}}, \label{eq: ansatz state}
\end{equation}
using an ansatz circuit (unitary operation)
\begin{equation}
    U(\bm{\theta}) \coloneqq U_l(\theta_l)\cdots U_2(\theta_2)U_1(\theta_1), 
\end{equation}
where $\bm{\theta} \coloneqq (\theta_1,\ldots,\theta_l)$ are real-valued parameters, $U_i(\theta_i)$ is a parametrized quantum gate, and $\ket{\psi_0}$ is some reference state.
In the VQE, we optimize these parameters so that the expectation value of the energy,
\begin{equation}
    L_{\mr{VQE}}(\bm{\theta}) \coloneqq \bra{\psi(\bm{\theta})}\hat{H}\ket{\psi(\bm{\theta})},
\end{equation}
is minimized with respect to $\bm{\theta}$.
For the evaluation of the expectation value $\bra{\psi(\bm{\theta})}\hat{H}\ket{\psi(\bm{\theta})}$, we usually consider the decomposition of the Hamiltonian into the sum of $n$-qubit Pauli operators; that is, we write as 
\begin{equation}
    \hat{H} \coloneqq \sum_i c_iP_i, 
\end{equation}
where $c_i$ is a real number and $P_i$ is a $n$-qubit Pauli operator. Then, the evaluation is reduced to the measurements of Pauli operators. 

Letting $\bm{\theta}^*$ be the optimum, we can regard $L_{\mr{VQE}}(\bm{\theta}^*)$ as a nice approximation of the ground-state energy and $\ket{\psi(\bm{\theta}^*)}$ as that of the ground state due to the variational principle.
During the process of the algorithm, the preparation of the ansatz state $\ket{\psi(\bm{\theta})}$ and the measurements necessary to compute $L_{\mr{VQE}}(\bm{\theta})$ are demonstrated by a quantum computer while the optimization is performed solely by a classical computer.
This separation of roles between classical and quantum computers alleviates the hardware requirement for quantum computers and makes the algorithm possible to run even on NISQ devices.

\subsection{Prior Work on Cost of VQE\label{subsec:VQE_cost}}
Here, we briefly review prior work investigating costs required to perform the VQE.

Although the VQE algorithm is a classical-quantum hybrid algorithm that is implementable using near-term quantum computers, it has been argued that the amount of resources needed to practically achieve sufficient accuracy is excessively large~\cite{Wecker2015}.
Indeed, three levels of iterations are in the VQE algorithm~\cite{Hamamura2020}, which may lead to the large number of required resources: 
\begin{enumerate}
    \item classically updating the parameters $\theta$,
    \item estimating the expectation value of the qubit Hamiltonian through its decomposition into a weighted sum of Pauli operators,
    \item evaluating the expectation value of a Pauli operator in the Hamiltonian by sampling. 
\end{enumerate}
The cost for the first and the most outer iteration is involved with the classical optimizer and the ansatz we choose~\cite{McClean2016}. 
For example, see Refs.~\cite{Wecker2015,ZhouLeo2020,Nakanishi2020,Parrish2019,Kubler2020,Arrasmith2020} for the investigation for improving classical optimizer, and see Refs.~\cite{Wecker2015,Lee2019,Gard2019,DallaireDemers2019,Ibe2020,Wiersema2020} for the investigation of ansatz.
For the second level of iteration, previous research has extensively investigated the reduction of number of Pauli measurements and the enhancement of efficiency of measurements~\cite{McClean2016,Hamamura2020, Jena2019,Verteletskyi2020,Yen2020,Izmaylov2020,Gokhale2020,Bravyi2017}. 
The third and the most inner iteration mainly depends on the statistical error to achieve desired accuracy. Reference~\cite{WangDaochen2019} provided a theoretical analysis to improve the number of samples required. 

Nevertheless, especially in the case of molecular dynamics, the total cost is still excessive due to the fact that we must perform the VQE at each step of the simulation. 

\section{Measurement on NISQ Device}
In this section, we review the basic setup and concepts of quantum measurement on a NISQ device. 

Consider an $n$-qubit quantum system. 
Suppose that we aim to measure the expectation value $\braket{\psi|H|\psi}$ of a Hermitian operator $H$ with respect to some given state $\ket{\psi}$. 
In a practical setup, $H$ is usually be a Hamiltonian of a given quantum system, and $\ket{\psi}$ is the resulting state of our ansatz circuit. 
On a NISQ device, the estimation of the expectation value can be done by measuring $\ket{\psi}$ in an appropriate Pauli basis. 
How many times we perform this measurement is called \textit{shot number}. 
In the following, we explain how this evaluation can be done in more detail. 

In the measurement procedure, we use the fact that the Hermitian operator can be decomposed into a sum of Pauli operators. 
The single-qubit Pauli operators are defined as 
\begin{align}
\sigma_0 &\coloneqq 
    \left(
    \begin{array}{cc}
    1 & 0 \\
    0 & 1
    \end{array}
    \right),  \\
\sigma_1 &\coloneqq 
    \left(
    \begin{array}{cc}
    0 & 1 \\
    1 & 0
    \end{array}
    \right),\\
\sigma_2 &\coloneqq 
    \left(
    \begin{array}{cc}
    0 & -i \\
    i & 0
    \end{array}
    \right), \\
\sigma_3 &\coloneqq 
    \left(
    \begin{array}{cc}
    1 & 0 \\
    0 & -1
    \end{array}
    \right). 
\end{align}
It is known that an $n$-qubit Hermitian operator can be expressed as a sum of $n$-qubit Pauli operators
\begin{equation}
    H = \sum_{i \in \{0,1,2,3\}^n} c_i P_i 
\end{equation}
with real coefficients $c_i$. 
Here, $i \in \{0,1,2,3\}^n$ is a string indicating an $n$-qubit Pauli operator; that is, if $i = k_1k_2\cdots k_n$, then $P_i = \sigma_{k_1}\otimes \sigma_{k_2}\otimes \cdots \sigma_{k_n}$. 
With this decomposition, the expectation value is written as 
\begin{equation}
    \braket{\psi|H|\psi} = \sum_{i \in \{0,1,2,3\}^n} c_i \braket{\psi|P_i|\psi}.  
\end{equation}
Thus, evaluating the expectation value of $\braket{\psi|P_i|\psi}$ for all $i$ and summing them up with weights $c_i$, we have the desired value. 

Now suppose that we want to compute the expectation value of $P_{k_1\cdots k_n} = \sigma_{k_1} \otimes \cdots \otimes \sigma_{k_n}$ with respect to $\ket{\psi}$.  
Let us represent $\ket{\psi}$ in the basis corresponding to $P_{k_1\cdots k_n}$; that is, 
\begin{equation}
    \ket{\psi} = \sum_{l_1\cdots l_n \in \{0,1\}^n} \alpha_{l_1\cdots l_n} \ket{l_1}\cdots \ket{l_n},
\end{equation}
where $\ket{l_j}$ is the Pauli-$\sigma_{k_j}$ basis. 
With this expression, it holds that 
\begin{equation}
\begin{aligned}
    &\braket{\psi|P_{k_1\cdots k_n}|\psi} \\
    &= \sum_{l_1\cdots l_n \in \{0,1\}^n} |\alpha_{l_1\cdots l_n}|^2 \braket{l_1|\sigma_{k_1}|l_1} \cdots \braket{l_n|\sigma_{k_n}|l_n}, 
\end{aligned}
\end{equation}
where $\braket{l_j|\sigma_{k_j}|l_j}$ takes either $1$ or $-1$ since $\ket{l_j}$ is an eigenstate of $\sigma_{k_j}$.  
Thus, we have that 
\begin{widetext}
    \begin{equation}
    \begin{aligned}
    \braket{\psi|P_{k_1\cdots k_n}|\psi}
    &= \sum_{\substack{l_1\cdots l_n \in \{0,1\}^n \\ \braket{l_1|\sigma_{k_1}|l_1} \cdots \braket{l_n|\sigma_{k_n}|l_n} = 1}} |\alpha_{l_1\cdots l_n}|^2 
    - 
    \sum_{\substack{l_1\cdots l_n \in \{0,1\}^n \\ \braket{l_1|\sigma_{k_1}|l_1} \cdots \braket{l_n|\sigma_{k_n}|l_n} = -1}} |\alpha_{l_1\cdots l_n}|^2.
    \end{aligned}
\end{equation}
\end{widetext}

Each $|\alpha_{l_1\cdots l_n}|^2$ can be evaluated by counting the results of the measurement of $\ket{\psi}$ in ${\ket{l_1}\cdots\ket{l_n}}$ basis. 
The total number of measurements we perform for the evaluation of $\braket{\psi|H|\psi}$ is called the shot number. 
The more times we perform this measurement, the more accurately we can estimate $|\alpha_{l_1\cdots l_n}|^2$.
That is, we can estimate $\ket{\psi|H|\psi}$ more accurately with a larger shot number, but we cannot exactly estimate it as long as the shot number is finite. 

\section{Derivation of Fluctuation-Dissipation Relation in Quantum Car-Parrinello Langevin Dynamics}~\label{sec:FDT}
In this section, we derive the fluctuation-dissipation theorem~\eqref{eq:FD_gamma} and \eqref{eq:FD_zeta} in our equations. 
Hereafter, we consider $1$-dimensional and $1$-parameter dynamics for brevity. One may easily extend our derivation to the general case. 

We first derive the Fokker-Planck equation of the probability distribution $\Phi(R,v,\theta,\xi,t)$ of the Langevin-type dynamics we have.
Let $G(R,v,\theta,\xi)$ be some statistics depending on $R$, $v$, $\theta$, and $\xi$. 
Then, we define 
\begin{equation}
    \begin{aligned}
        &\hat{G}(R(t),v(t),\theta(t),\xi(t)) \\
        &\coloneqq \int \mathrm{d}R\,\mathrm{d}v\,\mathrm{d}\theta\,\mathrm{d}\xi\,\, G(R,v,\theta,\xi)\Phi(R,v,\theta,\xi,t).
    \end{aligned}
\end{equation}
Consider the time derivative of $\hat{G}$. 
Then, on one hand, we have 
\begin{equation}~\label{derivative_time}
    \begin{aligned}
        \frac{\mathrm{d}\hat{G}}{\mathrm{d}t} = \int \mathrm{d}R\,\mathrm{d}v\,\mathrm{d}\theta\,\mathrm{d}\xi\,\, G(R,v,\theta,\xi)\frac{\partial\Phi(R,v,\theta,\xi,t)}{\partial t}.
    \end{aligned}
\end{equation}
On the other hand, it holds that
\begin{widetext}
    \begin{equation}~\label{derivative_coord}
\begin{aligned}
        \frac{\mathrm{d}\hat{G}}{\mathrm{d}t} 
        &= \frac{1}{\mathrm{d}t} \left[\hat{G}(R(t+\mathrm{d}t),v(t+\mathrm{d}t),\theta(t+\mathrm{d}t),\xi(t+\mathrm{d}t)) - \hat{G}(R(t),v(t),\theta(t),\xi(t)))\right]\\
        &=\frac{1}{\mathrm{d}t}\int \mathrm{d}R\,\mathrm{d}v\,\mathrm{d}\theta\,\mathrm{d}\xi\,\, 
        \left[
        \frac{\partial G}{\partial R}\mathrm{d}R
        + \frac{\partial G}{\partial v} \mathrm{d}v
        + \frac{\partial^2 G}{\partial v} (\mathrm{d}v)^2
        +\frac{\partial G}{\partial \theta} \mathrm{d}\theta
        + \frac{\partial G}{\partial \xi} \mathrm{d}\xi
        + \frac{\partial^2 G}{\partial \xi} (\mathrm{d}\xi)^2
        \right]\Phi \\
        &= \int \mathrm{d}R\,\mathrm{d}v\,\mathrm{d}\theta\,\mathrm{d}\xi\,\, 
        G\Bigg[
        -v\frac{\partial \Phi}{\partial R}
        - \frac{\partial}{\partial v} \left(\Phi\left(-\gamma v -\frac{1}{m}\frac{\partial L}{\partial R}\right)\right)
        + \frac{c}{2m^2\rho}\frac{\partial^2 \Phi}{\partial v^2}\\
        &\hspace{4cm}
        -\xi\frac{\partial \Phi}{\partial \theta}
        - \frac{\partial}{\partial \xi} \left(\Phi\left(-\zeta \xi -\frac{1}{\mu}\frac{\partial L}{\partial \theta}\right)\right)
        + \frac{c_{\theta}}{2\mu^2\rho}\frac{\partial^2 \Phi}{\partial \xi^2}
        \Bigg]. 
        \end{aligned}
\end{equation}
\end{widetext}
We used the Langevin equations~\eqref{Langevin_v}, \eqref{Langevin_R}, \eqref{Langevin_xi}, and \eqref{Langevin_theta}. 
Taking the limit of $\mathrm{d}t \to 0$, we ignore terms as small as $O(\sqrt{\mathrm{d}t})$.
Then, comparing Eqs.~\eqref{derivative_time} and \eqref{derivative_coord}, we have the Fokker-Planck equation
\begin{equation}~\label{FP_eq}
\begin{aligned}
    \frac{\partial\Phi}{\partial t} 
    &= -v\frac{\partial \Phi}{\partial R}
        - \frac{\partial}{\partial v} \left(\Phi\left(-\gamma v -\frac{1}{m}\frac{\partial L}{\partial R}\right)\right)
        + \frac{c}{2m^2\rho}\frac{\partial^2 \Phi}{\partial v^2}\\
        &\quad
        -\xi\frac{\partial \Phi}{\partial \theta}
        - \frac{\partial}{\partial \xi} \left(\Phi\left(-\zeta \xi -\frac{1}{\mu}\frac{\partial L}{\partial \theta}\right)\right)
        + \frac{c_{\theta}}{2\mu^2\rho}\frac{\partial^2 \Phi}{\partial \xi^2}. 
\end{aligned}
\end{equation}

Next, with this equation, consider the equilibrium distribution 
\begin{equation}
    \frac{\partial \Phi}{\partial t} = 0. 
\end{equation}
Then, assuming that the equilibrium distribution is given by a Boltzmann distribution defined in Eq.~\eqref{Eq:equib_dist}, we have 
\begin{widetext}
    \begin{equation}~\label{equib_eq}
    \begin{aligned}
        0&=
        \left(\beta\frac{\partial L}{\partial R} v\right)
        + \left(\gamma -\left(\gamma v + \frac{1}{m}\frac{\partial L}{\partial R}\right)mv\right)
        + \frac{c}{2m^2\rho}\left(-\beta m + \beta^2m^2v^2\right)\\
        &-
        \left(\beta\frac{\partial L}{\partial \theta} \xi\right)
        + \left(\zeta -\left(\zeta \xi + \frac{1}{\mu}\frac{\partial L}{\partial \theta}\right)\mu\xi\right)
        + \frac{c_{\theta}}{2\mu^2\rho}\left(-\beta \mu + \beta^2\mu^2\xi^2\right)
    \end{aligned}
\end{equation}
\end{widetext}
by substituting Eq.~\eqref{Eq:equib_dist} to Eq.~\eqref{FP_eq}.
Equation~\eqref{equib_eq} is satisfied when 
\begin{align}
    \label{eq:FD_gamma_1dim}
    \gamma &= \frac{\beta c}{2\rho m},\\
    \label{eq:FD_zeta_1dim}
    \zeta &= \frac{\beta c_{\theta}}{2\rho \mu}, 
\end{align}
which is the $1$-dimensional and $1$-parameter version of Eqs.~\eqref{eq:FD_gamma} and \eqref{eq:FD_zeta}.
In the general case, we obtain Eqs.~\eqref{eq:FD_gamma} and \eqref{eq:FD_zeta} by straightforwardly extending this argument with the assumption that the evaluation of each entry is independent of each other. 

\end{document}